\documentclass[
 aip,
 cha,
 amsmath,amssymb,
 reprint,
 twoside
]{revtex4-1}

\usepackage{dcolumn}
\usepackage{bm}

\usepackage{graphicx}
\usepackage{color}
\usepackage[absolute]{textpos}
\begin{document}
%


\title{Local observability of state variables and parameters in nonlinear 
modeling quantified by delay reconstruction}

\author{Ulrich Parlitz}
\email{ulrich.parlitz@ds.mpg.de}
\affiliation{Max Planck Institute for Dynamics and Self-Organization, Am Fa\ss berg 17,  37077 G\"ottingen, Germany }
\affiliation{Institute for Nonlinear Dynamics, Georg-August-Universit\"at G\"ottingen, Am Fa\ss berg 17,  37077 G\"ottingen, Germany }

\author{Jan Schumann-Bischoff}
\email{jan.schumann-bischoff@ds.mpg.de}
\affiliation{Max Planck Institute for Dynamics and Self-Organization, Am Fa\ss berg 17,  37077 G\"ottingen, Germany }
\affiliation{Institute for Nonlinear Dynamics, Georg-August-Universit\"at G\"ottingen, Am Fa\ss berg 17,  37077 G\"ottingen, Germany }

\author{Stefan Luther}   
\email{stefan.luther@ds.mpg.de}
\affiliation{Max Planck Institute for Dynamics and Self-Organization, Am Fa\ss berg 17,  37077 G\"ottingen, Germany }
\affiliation{Institute for Nonlinear Dynamics, Georg-August-Universit\"at G\"ottingen, Am Fa\ss berg 17,  37077 G\"ottingen, Germany }

\date{\today}

\begin{abstract}
Features of the Jacobian matrix of the delay coordinates map are exploited for quantifying the robustness and reliability 
of state and parameter estimations for a given dynamical model using an observed time series. Relevant concepts of this 
approach are introduced and illustrated for  discrete and continuous time systems employing a filtered H\'enon map and a
R\"ossler system.
  
\end{abstract}

\keywords{Observability, parameter estimation, nonlinear modelling}

\maketitle

\begin{quotation}
For many physical processes dynamical models (differential equations or iterated maps) 
are available but often not all of their variables and parameters are known or can be (easily) measured. 
In meteorology, for example, sophisticated large scale models exist, which have to be continuously 
adapted to the true temporal changes of temperatures, wind speed, humidity, and other relevant 
physical quantities. 
To obtain a model that ``follows'' reality, measured data have to be repeatedly incorporated into the model.
In  geosciences this procedure is called {\em data assimilation},  
but the task to track state variables and system parameters by means of estimation methods
occurs also in other fields of physics and applications. However, not all observables provide the information required 
to estimate a particular unknown quantity.  In this article, we consider this problem of \textit{observability} in the
context of chaotic dynamics where sensitive dependance on initial conditions complicates any estimation method.
A quantitative characterization of local observability employing delay coordinates is used to answer the question
where in state and parameter space estimation of a particular state variable or parameter is feasible and where not.
\end{quotation}

\section{Introduction}   \label{intro}
%
\begin{textblock}{180}(18,262)
\noindent \small{Copyright 2014 American Institute of Physics. This article may 
be downloaded for personal use only. Any other use requires prior permission of 
the author and the American Institute of Physics. The following article appeared 
in U. Parlitz \textit{et al.}, Chaos \textbf{24}, 024411 (2014) and may 
be found at \url{http://dx.doi.org/10.1063/1.4884344}.}
\end{textblock}
To describe and forecast dynamical processes in physics and many other fields 
of science mathematical 
models are used, like, for example,  ordinary differential equations (ODEs), partial differential equations (PDEs), or
iterated maps. Some of these models are derived from first principles while others are the result of a
general black-box modeling approach (e.g., based on neural networks). These models typically contain two kinds of 
variables  and parameters: those that can be directly measured or are known beforehand (e.g., fundamental 
physical constants) and others whose values are unknown and very difficult to access\cite{VTK04,RBKT10}. To estimate the latter 
estimation methods have been devised that aim at extracting the required information from the dynamics, here 
represented by the model equations and the experimentally observed dynamical evolution of the underlying process.
Different approaches for solving this dynamical estimation problem have been devised in the past, including
(nonlinear) observer or synchronization schemes \cite{PJK96,NM97,HLN01,GB08,ACJ08,SO09,SRL09,K05,FMG05,CK07,Amritkar09},
particle filters \cite{L10}, 
a path integral formalism \cite{A09,QA10}, or optimization based algorithms \cite{CGA08,B10,SBP11}. 

Before applying such an estimation  method one may ask whether the available time series (observable)
actually contains the required information to estimate a particular unknown value. In control theory this
 is called \textit{observability} problem and it can for linear systems of ODEs be analyzed and answered by means of the 
so-called observability matrix \cite{Sontag,NS90}. Using derivative coordinates this approach can be generalized 
for nonlinear continuous systems  \cite{HK77,Sontag, N82}.  For state estimation of chaotic systems Letellier, Aguirre 
and Maquet \cite{LAM05, AL05, LAM06, LA09, FBL12} considered  continuous dynamical systems
\begin{equation}  \label{contsyst}
  \dot {\mathbf x} = \mathbf {f} ( \mathbf{x}) 
\end{equation}
that generate some observed signal 
$s(t) = h( \mathbf{x}(t)) \in \mathbb{R}$ where $\mathbf{x} \in {\cal{U}} \subset {\mathbb R}^M  $
is the state of the system, ${\cal{U}}$ is a smooth submanifold of $\mathbb{R}^M$, 
and $h: \mathbb{R}^M \to \mathbb{R}$ denotes a measurement or observation function.
Consider now $D$-dimensional { \em derivative coordinates} 
\cite{Aeyels,Takens,SYC91,KS97,book_HDIA} of the 
observed signal $s(t)$ 
\begin{equation} \label{diffemb}
  \mathbf{y} = \left(s, \dot s, \ddot s, \ldots ,  s^{(D-1)}    \right)   = F( \mathbf{x}) \in \mathbb{R}^D
\end{equation}
where $s^{(k)}$ stands for the $k$-th temporal derivative of $s(t)$,
$D$ is the \textit{reconstruction dimension}, and   
$F$ is called   {\em derivative coordinates map} \cite{SYC91}.
If this map is (at least locally) invertible, then we can uniquely determine the full state vector 
$\mathbf{x}(t) \in {\cal{U}}$ from the signal $s(t)$ and its higher derivatives \cite{LAF09}. Furthermore, small perturbations
in $\mathbf{y}$ should correspond to small perturbations in $\mathbf{x}$ and vice versa.
Therefore,  we want the map 
$F:  {\cal{U}}  \to   F({\cal{U}})  \subset \mathbb{R}^D$ to be an 
\textit{immersion}, i.e.  a smooth map whose derivative map is one-to-one at every point of ${\cal{U}}$,
or equivalently, whose Jacobian matrix $DF(\mathbf{x})$ has full rank on the tangent space 
(here $\mbox{rank}(DF(\mathbf{x})) = M \ \forall \mathbf{x} \in {\cal{U}}$).
This does \textit{not} imply that the map $F$ itself is 
one-to-one ($F(\mathbf{x}) = F(\mathbf{z}) \Rightarrow \mathbf{x} = \mathbf{z}$), since the
derivative coordinates 
\eqref{diffemb} may provide states $\mathbf{y} \in \mathbb{R}^D$ with two (or more) 
pre-images separated by a \textit{finite} distance.
Therefore, the observability analysis presented in the following is \textit{local}, only, because it 
is based on analyzing (the rank) of the Jacobian matrix $DF$. The (global)  one-to-one property 
of the map $F$ is \textit{not} checked (what would be necessary, and for compact U also sufficient,  to
show that $F$ is an \textit{embedding} \cite{SYC91}).

The  $D \times M$ Jacobian matrix $DF(\mathbf{x}) $ can be computed by means of the vector field given in 
Eq.~(\ref{contsyst}).  In fact, for linear ODEs the Jacobian matrix 
$DF(\mathbf{x})$  conforms with the \textit{observability matrix}  known 
from (linear) control  theory \cite{AL05}. 
To estimate the rank of $DF(\mathbf{x})$ Letellier, Aguirre, and Maquet \cite{LAM05,AL05} 
suggest to compute the  eigenvalues $\mu_k \ge 0$ of the $M \times M$ - matrix  
\begin{equation}
A(\mathbf{x}) = DF^{tr}(\mathbf{x}) \cdot DF(\mathbf{x}).  
\end{equation}
Nonzero eigenvalues indicate full rank of $DF(\mathbf{x})$ and thus 
local invertibility of $F$ at $\mathbf{x}$.
To quantify the (local) invertibility of $F(\mathbf{x})$ and thus the (local) observability of 
the full state $\mathbf{x}$ Aguirre, Letellier, and Maquet \cite{LAM05,AL05} 
introduced the {\em observability index} 
\begin{equation}  \label{obsindx}
   \delta(\mathbf{x}) = \frac{ \mu_{min} (A) }  { \mu_{max} ( A) }
\end{equation}
where $\mu_{min}(A) $ and $\mu_{max}(A)$ denote the smallest and 
the largest eigenvalue of the matrix $A$, respectively.
Time averaging (along the available trajectory for $0 \le t \le T$) yields
\begin{equation}  \label{averobsindx}
  \bar  \delta = \frac{ 1 }  { T }    \int_{0}^T \delta(\mathbf{x}(t)) dt   .
\end{equation}

Instead of derivative coordinates 
we consider in the following delay coordinates \cite{Aeyels,Takens,SYC91,KS97,book_HDIA}.
Furthermore, we extend the observability analysis 
to parameter estimation and compute a specific measure of uncertainty \cite{PSBL} for each 
state variable or parameter to be estimated. Last not least, we are not only interested in 
quantifying the average observability (like $ \bar  \delta$ in Eq.~(\ref{averobsindx})) but 
also in local variations that can be exploited during the state and parameter estimation process.

\section{Delay coordinates and observability}
To motivate the concepts to be presented in the following 
we shall first consider a discrete time system (iterated map)
where all model parameters 
are known and only state variables have to be estimated from
the observed time series.

\subsection{Estimating state variables of a filtered H\'enon map}
For an $M$ dimensional discrete system
\begin{equation}  \label{discrsyst}
   \mathbf{x}(n+1) = \mathbf{g} ( \mathbf{x}(n)) 
\end{equation}
which generates the times series $\{ s(n) \}$ with $s(n) = h( \mathbf{x}(n))$ 
where $n = 1, ..., N$ we can construct $D$ dimensional
{\em delay coordinates} \cite{Aeyels,Takens,SYC91,KS97,book_HDIA}
with reconstructed states
\begin{eqnarray} \label{delembforward}
  \mathbf{y}(n)  & = &  \left(s(n), s(n+1), ....,  s(n+D-1)    \right)  \\ \nonumber
                      &  = & G_+( \mathbf{x}(n)) \in \mathbb{R}^D.
\end{eqnarray}
Again we assume that all states of interest $\mathbf{x}$  lie within a  smooth manifold 
${\cal{U}} \subset \mathbb{R}^M$. Here we consider delay coordinates \textit{forward} in time. 
The function $G_+$ is therefore called  {\em forward delay coordinates map}  
$G_+:   {\cal{U}} \to  G_+({\cal{U}}) \subset  \mathbb{R}^D$. It is also possible
to use delay coordinates \textit{backward} in time, or mixed forward and backward, 
and we shall address this issue in Sec.~\ref{sec:fwemb}. 

As already discussed with  derivative coordinates in the previous section 
a state $\mathbf{x} = (x_1, \ldots, x_M)$ is locally observable from the time series $\{ s(n) \}$
if $G$ is an immersion, i.e. if the Jacobian matrix $DG(\mathbf{x})$ has maximal (full) rank $M$
at $\mathbf{x}$. The  corresponding $D \times M$ Jacobian matrix $DG(\mathbf{x})$ 
can be computed  using the iterated map Eq.~(\ref{discrsyst}).  
If the Jacobian matrix $DG(\mathbf{x})$ has maximal rank $M$ (assuming $M \le D$)
then $G$ is locally invertible (on $G({\cal{U}})$). ``Local''  means that still a delay vector $\mathbf{y}$ 
could possess different pre images (separated by a finite distance). 

To motivate and illustrate this analysis we consider the H\'enon map
\begin{eqnarray} \label{Henon}  
x_1(n+1) & = & 1 -  a  x_1^2(n) + b x_2(n)  \\
x_2(n+1) & = & x_1(n)
\end{eqnarray}
with parameters $a=1.4$ and $b=0.3$.
In the following we shall assume that the dynamics of this system is 
observed via a filtered signal $s(n)$ provided by an FIR-filter
\begin{eqnarray} \nonumber
s(n) & = & x_1(n) + c x_1(n-1)  \\
       & = & x_1(n) + c x_2(n) = h( \mathbf{x}(n))   \label{FIR}
       \end{eqnarray}
with filter parameter $c$.

\begin{figure} [!ht]
\centering

\includegraphics[width=8.7cm]  {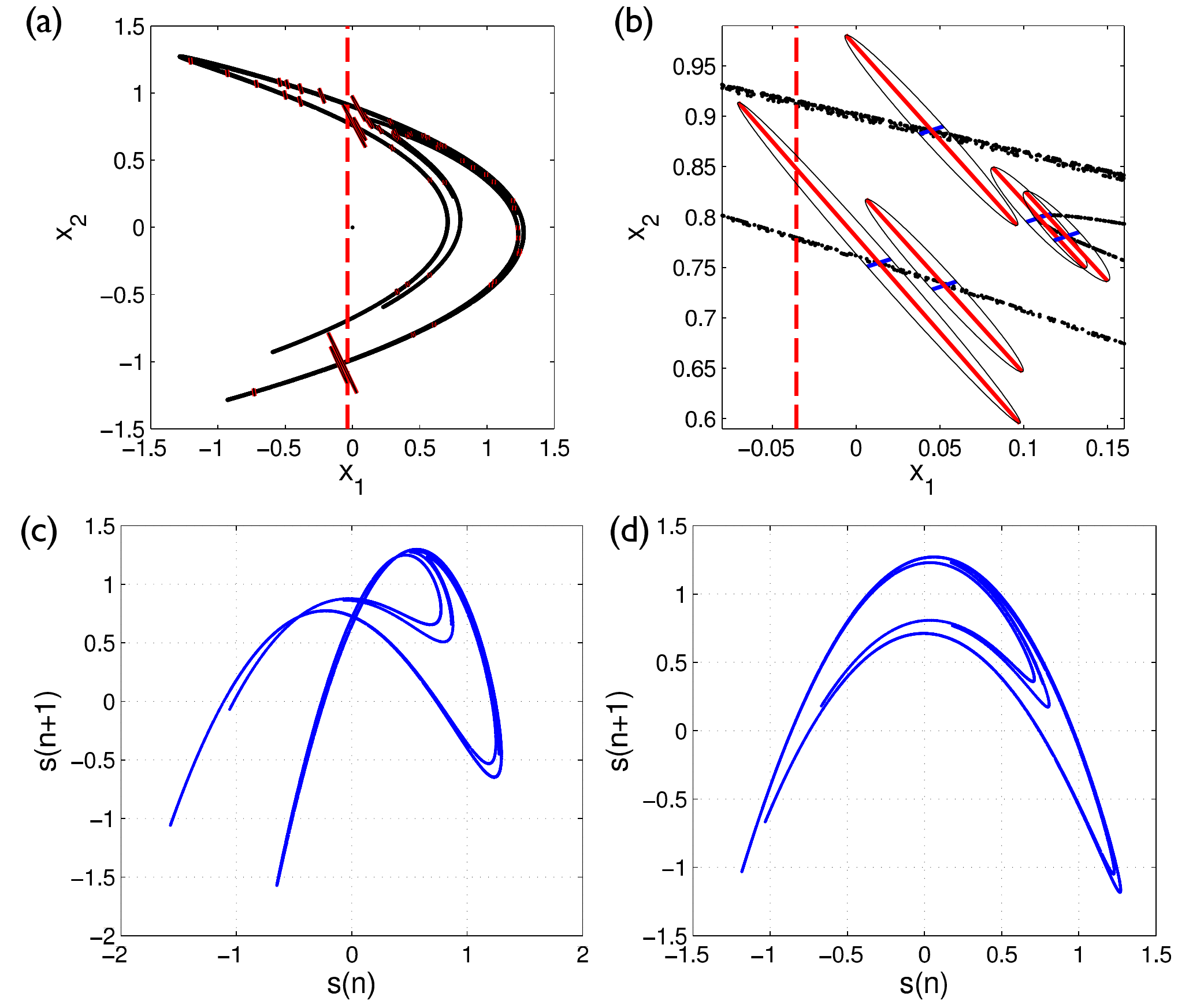}
\caption{(Color online) (a), (b) H\'enon attractor and singular axes $\frac{1}{\sigma_1} \mathbf{V}^{(1)}$ (short, blue) 
and $\frac{1} {\sigma_2} \mathbf{V^{(2)}}  $  (longer, red) for filter parameter  $c=0.5$.
(c), (d) Delay coordinates for $c=0.5$ ($x_1^s = -0.0357$) and $c=0.08$ ($x_1^s = -1.311$). }
\label{fig1}
\end{figure}

For two dimensional delay coordinates the delay coordinates map reads
\begin{eqnarray*}
        G({\bf x}(n)) & = & ( s(n), s(n+1) ) \\
                          & = & (x_1(n) + c x_2(n),  \\  
                          &    &  \  \   \  \  \  \  \  \   1-a x_1^2(n) +  b x_2(n) + c x_1(n) ) 
\end{eqnarray*}
or 
\begin{equation}
    G(\mathbf{x}) =   (x_1 + c x_2,  1-a x_1^2 +  b x_2 + c x_1 )  .
\end{equation}
The Jacobian matrix of the map $G$ is given by:
\begin{equation}
  DG(\mathbf{x}) = \left(   \begin{array} {ccc}
                                         1                     &    c  \\
                                            -2ax_1 + c    &    b
                                \end{array} \right)
\end{equation}
and its determinant 
\begin{equation}
    \det ( DG( \mathbf{x} ) ) =  2ac x_1 +b - c^2
\end{equation}
vanishes for all states $\mathbf{x} = (x_1, x_2)$ on the singular line
\begin{equation} \label{singular}
   x_1^s = \frac{c^2 - b} {2ac} .
\end{equation}
For $c \to 0$ the FIR-filter is (asymptotically) deactivated and the critical line
disappears ($c \to 0 \Rightarrow x_1^s \to - \infty$). For $0.0867 < c < 3.66$, however,
the critical line crosses the chaotic attractor as shown for $c=0.5$ in Fig.~\ref{fig1}a.

\subsection{Forward, backward, and mixed  delay coordinates} \label{sec:fwemb}
Instead of using state space reconstruction based on forward delay coordinates \eqref{delembforward}
one could also use \textit{backward} delay coordinates 
\begin{eqnarray} \label{delembbackward}
  \mathbf{y}(n)  & = &  \left(s(n), s(n-1), \ldots ,  s(n-D+1)    \right)  \\ \nonumber
                      &  = & G_{-}( \mathbf{x}(n)) \in \mathbb{R}^D
\end{eqnarray}
or more general, a combination of forward and backward components
\begin{eqnarray} \nonumber
  \mathbf{y}(n)  & = &  \left(s(n-D_{-}), \ldots s(n-1), s(n), s(n+1), \right.  \\  \label{delembmixed}
                        &   &   \left.   \ldots,  s(n+D_{+})    \right)  \\ \nonumber
                      &  = & G_{\pm}( \mathbf{x}(n); D_{-}, D_{+}) \in \mathbb{R}^D
\end{eqnarray}
called \textit{mixed} delay coordinates in the following, with reconstruction dimension
$D=1+D_{-}+D_{+}$. 
To obtain the backward components $s(n-k) = h( ( {\mathbf x}(n-k))$ the 
inverse map $ {\mathbf x}(n-1) = {\mathbf g}^{-1} ( {\mathbf x}(n)) $ 
and its Jacobian matrix are required (here we assume that the dynamics is time invertible).
For discrete time systems (like the H\'enon example) 
the underlying map \eqref{discrsyst} has to be inverted 
and for continuous time systems the inverse of the flow can in principle be computed 
by integrating the system ODEs \eqref{contsyst} backward in time. In both cases, however,
problems may occur in practice, because an explicit form of the inverse map may not exist  and
backward integration of dissipative systems results in diverging solutions and numerical 
instabilities (for longer integration times). 
Despite these difficulties inclusion of backward components turns out to be beneficial 
for the estimation task as will be demonstrated in the following for the H\'enon examples and 
the R\"ossler system.

\subsection{Noisy observations and uncertainty}
At states $(x_1,x_2)$  with  $x_1 \ne x_1^c$ the delay coordinates map  $G$ is in principle 
invertible, but the inverse can be very susceptible to perturbations in $\mathbf{y}$ like measurement noise.
To quantify the robustness and the sensitivity of the inverse with respect to noise we 
consider the singular value decomposition of the Jacobian matrix $DG$ of the delay coordinates map
\begin{equation} \label{SVD1}
   DG = U \cdot S \cdot V^{tr}
\end{equation}
where $S = \rm{diag}( \sigma_1, \ldots , \sigma_M) $ is an $M \times M$ diagonal matrix containing the 
singular values $\sigma_1 \ge \sigma_2 \ge \ldots \ge \sigma_M \ge 0$  and $U = (\mathbf{u}^{(1)}, \ldots , 
\mathbf{u}^{(M)}) $ and $V = ( \mathbf{v}^{(1)}, \ldots , \mathbf{v}^{(M)} ) $ are orthogonal matrices, represented by the
column vectors $\mathbf{u}^{(i)} \in \mathbb{R}^D$ and $\mathbf{v}^{(i)} \in \mathbb{R}^M$, respectively.
$V^{tr}$ is the transposed of $V$ coinciding with the inverse $V^{-1} = V^{tr}$. Analogously,
$U^{tr} = U^{-1} $ and the inverse Jacobian matrix reads
\begin{equation} \label{SVD2}
   DG^{-1} =  V \cdot S^{-1} \cdot U^{tr} 
\end{equation}
where $S^{-1} = \rm{diag}(1/\sigma_1, \ldots, 1/ \sigma_M) $.
Multiplying by $U$ from the right we obtain $ DG^{-1} U =  V \cdot S^{-1}  $ or
\begin{equation} \label{SVD3}
   DG^{-1}  {\mathbf{u}}^{(m)} = \frac{1}{\sigma_m} {\mathbf{v}}^{(m)}  \  \  \  \  \  (m = 1, \ldots, M) .
\end{equation}

\begin{figure} 
\centering
\includegraphics[width=8.5cm]   {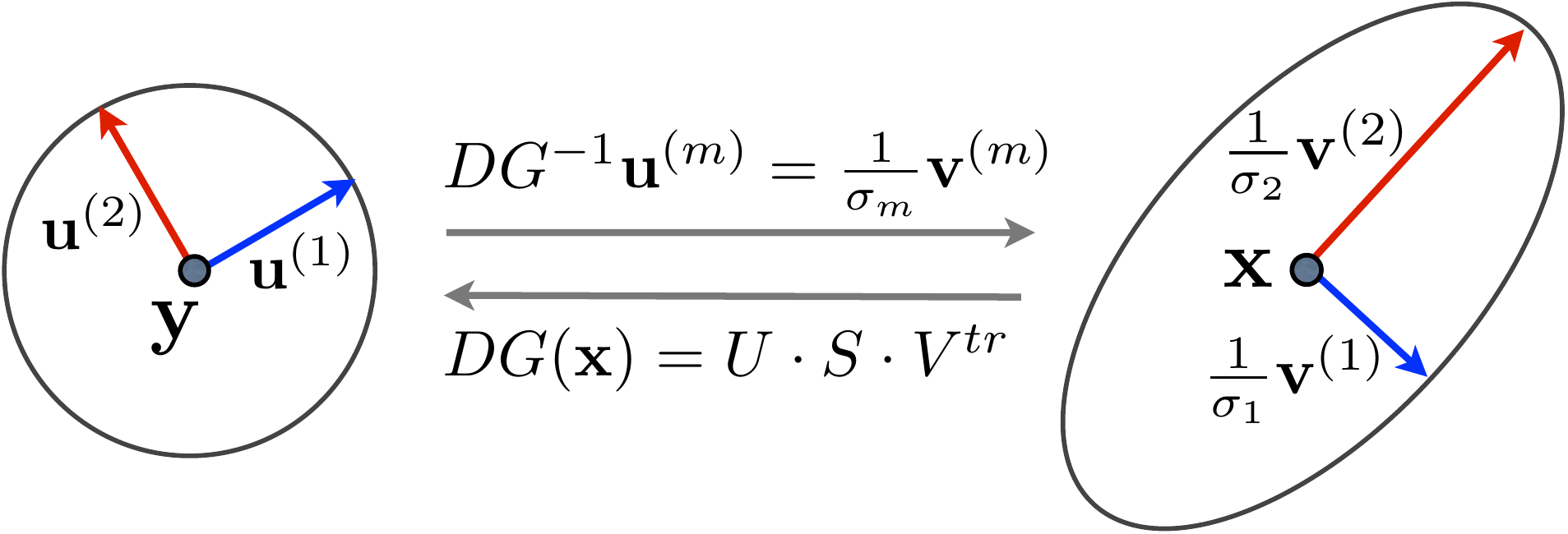}
\caption{(Color online) The inverse Jacobian $DG^{-1}(\mathbf{y})$ maps perturbations of $\mathbf{y}$ in delay reconstruction space 
to deviations from the state $\mathbf{x}$ whose magnitudes depend on the direction of the perturbation as 
described by Eq.~(\ref{SVD3}). }
\label{fig2}
\end{figure}

\begin{figure} 
\centering
 \includegraphics[width=8.7cm] {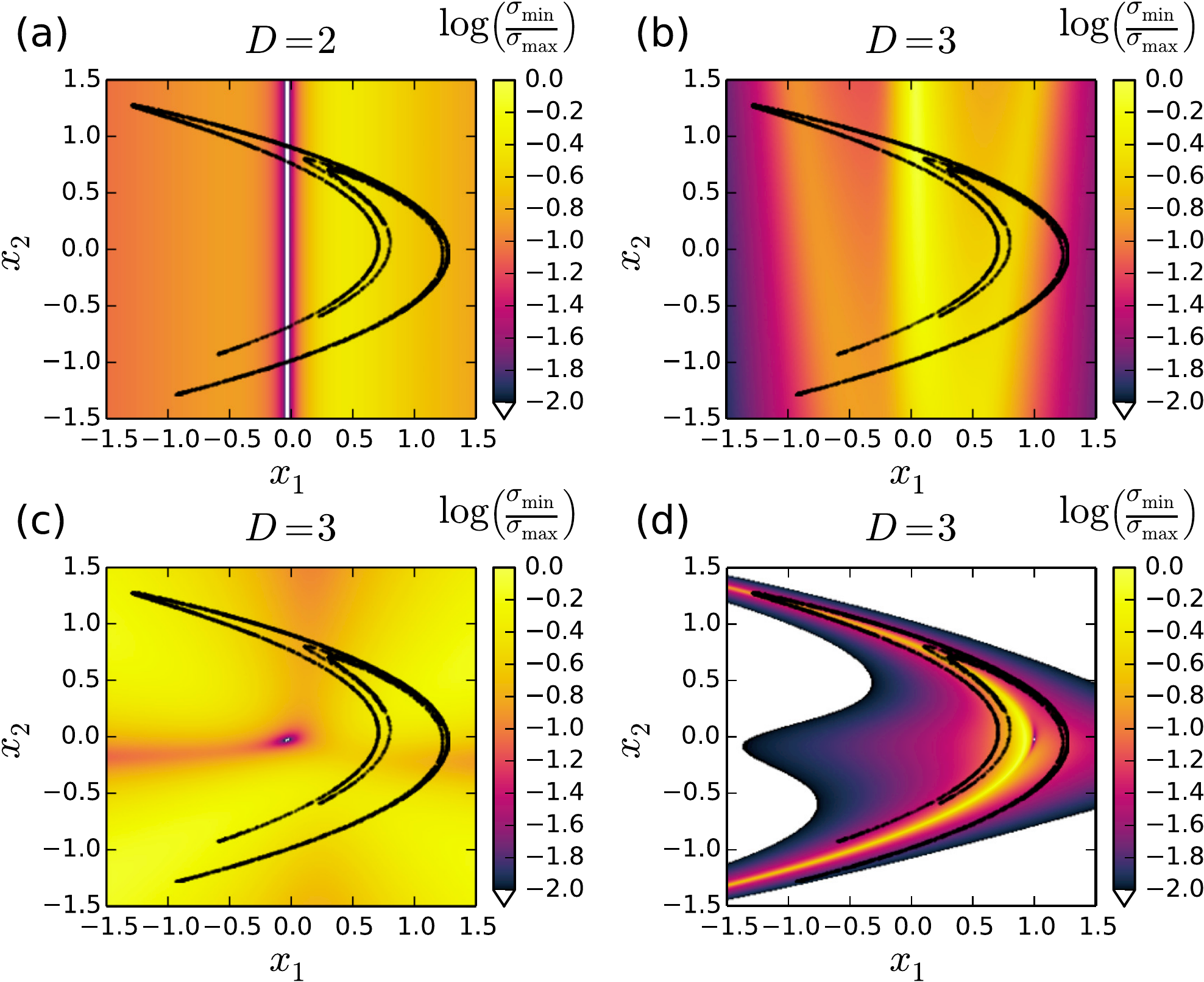} 
  \caption{(Color online) Local observability of the filtered H\'enon map \eqref{Henon}-\eqref{FIR}.
               Logarithm of the (color coded) ratio of the smallest singular value $\sigma_{min}= \sigma_M$ ($M=2$) divided 
              by largest singular value $\sigma_{max} = \sigma_1$ 
             vs. coordinates $x_1$ and $x_2$  for $c = 0.5$. 
             (a) $D=2$ dimensional forward delay coordinates \eqref{delembforward},
             (b) $D=3$ dimensional forward delay coordinates \eqref{delembforward},
             (c) $D=3$ dimensional mixed delay coordinates \eqref{delembmixed} ($D_{-} = 1 = D_{+}$), and 
             (d) $D=3$ dimensional backward delay coordinates \eqref{delembbackward}.} 
\label{fig3}
\end{figure}

This transformation is illustrated in Fig.~\ref{fig2} and it describes how small perturbations of $\mathbf{y}$
in delay reconstruction space result in deviations from $\mathbf{x}$ in the original state space. 
Most relevant for the observability of the (original) state $\mathbf{x}$ is the length of the longest principal axis 
of the ellipsoid given by the inverse of the smallest singular value $\sigma_M$ (see Fig.~\ref{fig2}). 
Small singular values correspond to directions in state space where it is difficult (or even impossible) to 
locate the true state $\mathbf{x}$ given a finite precision of the reconstructed state $\mathbf{y}$. 
For the filtered H\'enon map we find that  the closer the 
state $\mathbf{x}$ is to the critical line given by $x_1^s$ \eqref{singular} the more severe is this uncertainty. 
This is illustrated in Fig.~\ref{fig1}a,b
where at some points $\mathbf{x}$ the ellipses spanned by the column vectors of the matrix $V\cdot S^{-1}$ 
are plotted.
Figure~\ref{fig3} shows (color coded) the logarithm of the ratio smallest singular value $\sigma_{min} = \sigma_M$ 
(here: $M=2$) divided by the
largest singular value $\sigma_{max} =\sigma_1$  vs. state variables $x_1$ and $x_2$ 
in a range of coordinates containing the chaotic H\'enon attractor. In Fig.~\ref{fig3}a  $D=2$ dimensional
forward delay \eqref{delembforward} is considered where at $x_1^s$  the smallest 
singular value $\sigma_{min}=\sigma_{M} = \sigma_2$ vanishes indicating the 
singularity (\ref{singular})  illustrated in Fig.~\ref{fig1}a,b. If the reconstruction dimension $D$ is increased from $D=2$ 
to $D=3$  the singularity disappears as can be seen in Fig.~\ref{fig3}b showing the ratio $\sigma_{min}/\sigma_{max}$ 
(color coded) for $D=3$ dimensional forward delay coordinates. For comparison, Figs.~\ref{fig3}c,d show results 
obtained with mixed delay coordinates \eqref{delembmixed}  and backward delay coordinates \eqref{delembbackward},
respectively. The white areas in Fig.~\ref{fig3}d correspond to ratios $\sigma_{min} / \sigma_{max} < 0.01$ indicating poor 
observability (due to fast divergence of backward iterates of the H\'enon map).
Further increase of the reconstruction dimension ($D=4$ or $D=5$) 
results in even larger values of $\sigma_{min} / \sigma_{max}$ (not shown here).

\begin{figure} 
\centering
 \includegraphics[width=8.7cm] {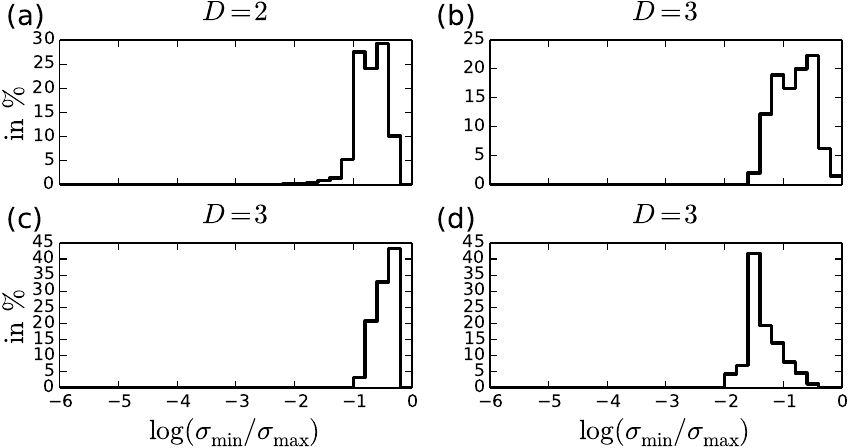}
  \caption{(Histograms of rations $\sigma_{min} / \sigma_{max}$ 
  computed at $10^6$ points of the H\'enon attractor with:
      (a) $D=2$ dimensional forward delay coordinates \eqref{delembforward},
             (b) $D=3$ dimensional forward delay coordinates \eqref{delembforward},
             (c) $D=3$ dimensional mixed delay coordinates \eqref{delembmixed} ($D_{-} = 1 = D_{+}$), and 
             (d) $D=3$ dimensional backward delay coordinates \eqref{delembbackward}.
             Compare corresponding diagrams in Fig.~\ref{fig3}. }
\label{fig4}
\end{figure}

To assess the observability on the H\'enon attractor we computed 
histograms of ratios $\sigma_{min} / \sigma_{max}$ at $10^6$ points. Figure~\ref{fig4} shows
these histograms for the same coordinates used to generate the corresponding diagrams in Fig.~\ref{fig3}.
The best results (large ratios) provide  mixed delay coordinates (Fig.~\ref{fig4}c). We speculate that this
is due to the fact that  forward and backward components cover different directions in state space (similar to 
Lyapunov vectors corresponding to positive and negative Lyapunov exponents).

%
If the perturbations of the reconstructed state $\mathbf{y}$ are due to normally distributed measurement noise they can be described
by a symmetric Gaussian distribution centered at $\mathbf{y}$
\begin{equation}
    Q(\mathbf{\tilde y}) = \frac{ \exp \left[ -\frac{1}{2} ({\mathbf{\tilde y}} - {\mathbf{y}})^{tr} \Sigma_y^{-1} ({\mathbf{\tilde y}} - {\mathbf{y}}) \right]
   }{ \sqrt{  (2\pi)^D \det (\Sigma_y)  }   }
\end{equation}
where $\Sigma_y = {\rm{diag}}(\rho^2, \ldots, \rho^2) = \rho^2 I_D $ denotes  the $D \times D$ covariance matrix
($I_D$ stands for the $D$-dimensional unit matrix) and the standard deviation $\rho$ quantifies the noise amplitude.
For (infinitesimally) small perturbations $\mathbf{\Delta y}= \mathbf{\tilde y} - \mathbf{y} $ this distribution is 
mapped by the (pseudo) inverse of the delay coordinates map to the (non-symmetrical) distribution 
\begin{equation}
   P(\mathbf{\tilde x}) = \frac{\exp \left[ -\frac{1}{2} (\mathbf{\tilde x} - \mathbf{x})^{tr} \Sigma_x^{-1} (\mathbf{\tilde x} - \mathbf{x}) \right]}{ \sqrt{  (2\pi)^M \det (\Sigma_x)  } } 
    \end{equation}
centered at $\mathbf{x}$ with the inverse covariance matrix
\begin{eqnarray} \label{sigma_x_inv}
   \Sigma_x^{-1}  & = &  DG^{tr} \cdot \Sigma_y^{-1} \cdot DG  = \frac{1}{\rho^2}  DG^{tr} \cdot DG \\
                         & = &  \frac{1}{\rho^2} V \cdot S^2 \cdot V^{tr} . 
\end{eqnarray} 
The marginal distribution $P_j$ of a given state variable $\tilde x_j$ 
centered at $x_j$ is given by
\begin{equation}
  P_j( \tilde  x_j) = \frac{1}{\rho_j  \sqrt{2 \pi} }  \exp \left[ - \frac{(\tilde x_j - x_j)^2} {2 \rho_j^2}  \right]
\end{equation}
where the standard deviation $\rho_j$ is given by the square root of the diagonal elements of the covariance matrix
\begin{equation}
    \rho_j = \sqrt{ \Sigma_{x,jj} }  .
\end{equation}
Using Eq.~(\ref{sigma_x_inv}) the standard deviation of the marginal distribution $P_j$ can be written
\begin{equation}
    \rho_j = \rho  \sqrt{ \left[ DG^{tr} \cdot DG  \right]^{-1}_{jj}} = \rho  \sqrt{\left[ V \cdot S^{-2} \cdot V^{tr} \right]_{jj}}
\end{equation}
and we consider in the following the factor

\begin{equation} \label{uncert}
   \nu_j = \sqrt{ \left[ DG^{tr} \cdot DG  \right]^{-1}_{jj}} =  \sqrt{\left[ V \cdot S^{-2} \cdot V^{tr} \right]_{jj}}
\end{equation}
as our  measure of \textit{uncertainty} when estimating $x_j$, because it quantitatively describes how the 
initial standard deviation $\rho$ is amplified when estimating variable $x_j$ \cite{PSBL}.

\begin{figure} [!ht]
\centering
 \includegraphics[width=8.7cm]   {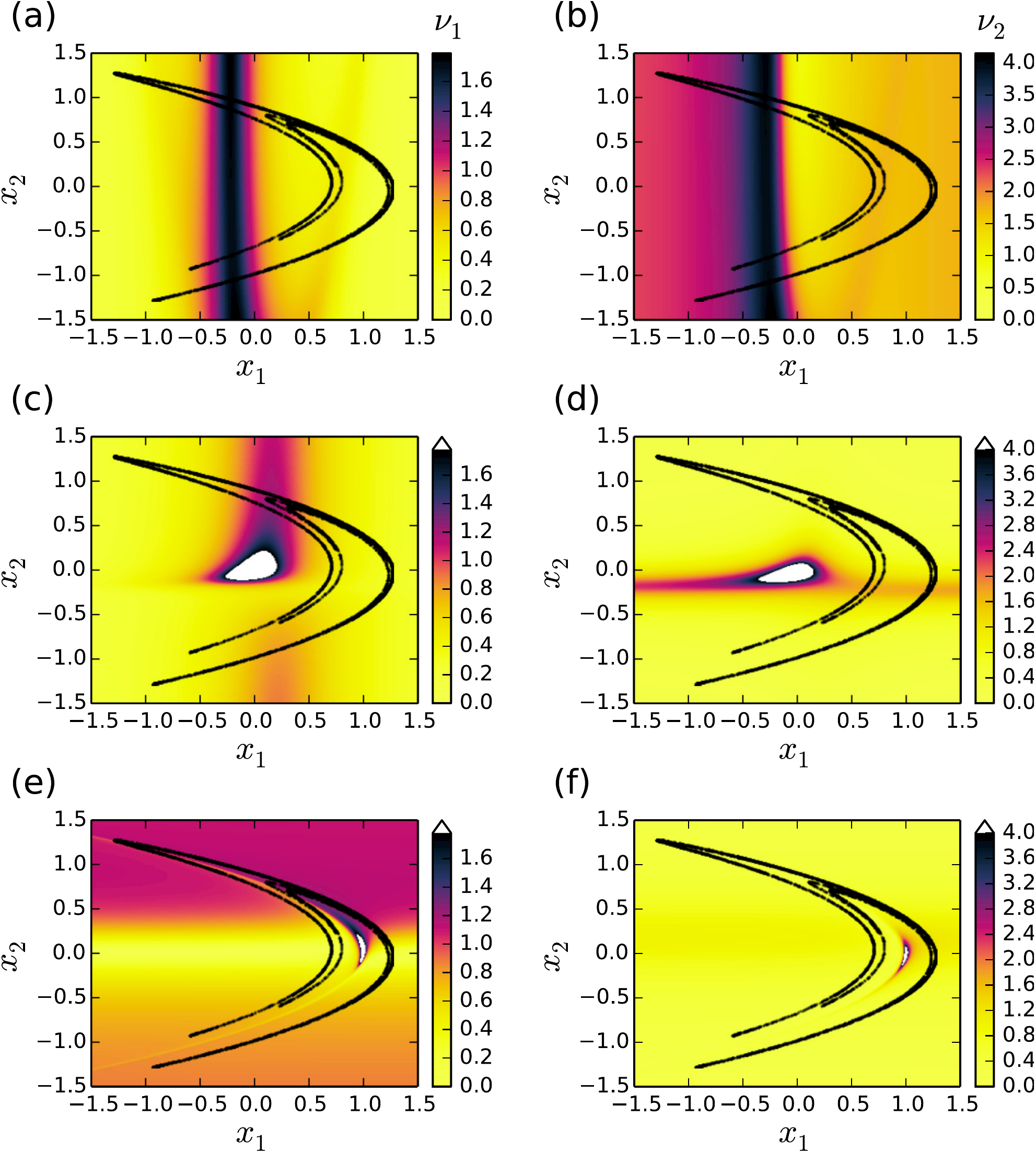}  %
   \caption{(Color online) Uncertainty (\ref{uncert}) of the variables $x_1$ and $x_2$ of the H\'enon map for $c=0.5$ and 
 different $D=3$ dimensional delay coordinates.
(a), (b) forward coordinates \eqref{delembforward}, (c), (d) mixed coordinates  \eqref{delembmixed}, and (e), (f) backward coordinates  \eqref{delembbackward}.}
\label{fig5}
\end{figure}

\begin{figure} [!ht]
\centering
 \includegraphics[width=8.7cm]   {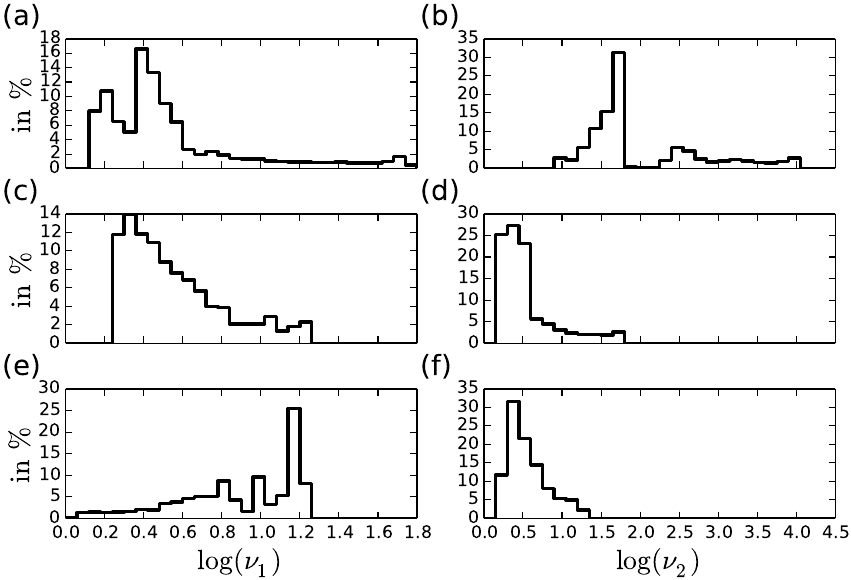}
  \caption{Histograms of uncertainties (\ref{uncert}) of $x_1$ and $x_2$ on the H\'enon attractor (computed at $10^6$ points) for $D=3$ and $c=0.5$ with (a), (b) forward, (c), (d) mixed, and (e), (f) backward delay coordinates (comp. Fig.~\ref{fig4}).}
\label{fig6}
\end{figure}

Figure~\ref{fig5} shows the uncertainties $\nu_1$ and $\nu_2$ vs. $(x_1, x_2)$ for different $D=3$ dimensional 
delay coordinates. In Fig.~\ref{fig4}a,b results obtained with $D=3$ dimensional
forward delay coordinates \eqref{delembforward} are given.
Large uncertainties occur mainly in a vertical stripe located near the singularity at $x_1^s$ (Eq.~(\ref{singular})) occurring 
for $D=2$. Figures~\ref{fig5}c,d,e,f show uncertainties of $x_1$ and $x_2$ obtained with mixed delay coordinates ((c),(d))
and backward delay coordinates ((e), (f)). For mixed delay coordinates (Fig.~\ref{fig5}c,d) areas with very high uncertainties 
occur near the origin, but along the attractor  $\nu_1$ and $\nu_2$ take only relatively low values.
This is also confirmed by the $\nu$-histograms on the attractor given in Fig.~\ref{fig6} for the same delay coordinates as 
used in Fig.~\ref{fig4}. Again, the mixed delay coordinates turns out to be  superior to the purely forward or backward coordinates. Furthermore, 
the dependance of the range of uncertainty values on the type of coordinates is different for different variables. While the uncertainty 
$\nu_1$ of $x_1$ increases
when changing from forward to backward delay coordinates (Figs.~\ref{fig6}a,e), the uncertainty of $x_2$ exhibits the opposite 
trend (Figs.~\ref{fig6}b,f). 

%

\subsection{State and  parameter estimation} \label{sec:bfemb}

Until now only the state variables $x_1$ and $x_2$ are considered as unknowns to be estimated
and the parameters $a$ and $b$ of the H\'enon map and $c$ of the FIR filter are assumed to be
known. We shall now consider the general case including unknown parameters 
$\mathbf{p} = (p_1, \ldots, p_P) \in \mathbb{R}^P$ of the dynamical system and unknown parameters
$\mathbf{q} = (q_1, \ldots, q_Q) \in \mathbb{R}^Q$  of the measurement function $h(\mathbf{x}, \mathbf{q})$.
Let the dynamical model be a $M$-dimensional discrete
\begin{equation} \label{discrsyst2}
     \mathbf{x}(n+1) = {\mathbf g} (\mathbf{x}(n), \mathbf{p})
\end{equation}
or a continuous
\begin{equation}  \label{contsyst2}
  \dot {\mathbf x} = {\mathbf f} ( {\mathbf x}, \mathbf{p}) 
\end{equation}
dynamical system generating a flow
\begin{equation} \label{flow}
  \phi^t : \mathbb{R}^M \rightarrow \mathbb{R}^M
\end{equation}
with discrete $t=n \in \mathbb{Z} $ or continuous $t \in \mathbb{R} $ time.
Furthermore, let's assume that a time series $\{ s (n) \}$ of length $N$ is given
observed via the measurement function
\begin{equation}
      s(t) = h( \phi^{t}(\mathbf{x}, \mathbf{p}) , \mathbf{q}) 
\end{equation}
from a trajectory starting at $\mathbf{x}$.

This provides the $D$-dimensional  delay coordinates 
\begin{eqnarray} \nonumber
   \mathbf{y} & = & \left(s(-D_{-}\tau_{-}), \ldots , s(-\tau_{-}), s(0),  s(\tau_{+}), \ldots,  s(D_{+}\tau_{+})    \right)  \\ \label{mixit}
                  &  = & G( \mathbf{x}, \mathbf{p}, \mathbf{q} ; D_{-},D_{+},\tau_{-}, \tau_{+}) \in \mathbb{R}^D
\end{eqnarray}
with $D = 1+ D_{-} + D_{+}$.
Here the delay coordinates map $G$ is considered as a function of: (i) the state $\mathbf{x}$ and the parameters
$\mathbf{p}$ of the underlying system,  (ii) the parameters $\mathbf{q}$ of the measurement function, 
(iii) the dimension parameters $D_{-}$ and $D_{+}$, and (iv) the delay times $\tau_{-}$ and $\tau_{+}$ in
backward and forward direction, respectively. 
The option to use different delay times, $\tau_{-}$ and $\tau_{+}$ for the backward and forward iterations is motivated 
by the fact that for dissipative systems backward solutions $\phi^{\tau_{-}}(\mathbf{x}) $ quickly diverge and therefore
a choice $\tau_{-} < \tau_{+}$ may be more appropriate. For the same reason $D_{-}$ has typically to be smaller than $D_{+}$. 
Since the reconstruction dimensions and the delay times are chosen a priori and are not part of the estimation problem
they shall not be listed as arguments of $G$ to avoid clumsy notation.
The Jacobian matrix $DG( \mathbf{x}, \mathbf{p}, \mathbf{q})$ of $G$ has the structure
\begin{equation}
  DG(\mathbf{x}, \mathbf{p}, \mathbf{q}) = (A, B ,C)
\end{equation}
where:
\begin{equation*} 
    A = \left(    \begin{array}  {ccc}
          \frac{\partial h (\phi^{-D_{-}\tau_{-}}(\mathbf{x},\mathbf{p}), \mathbf{q}) } {\partial x_1}  &  \ldots  &  
          \frac{\partial h (\phi^{-D_{-}\tau_{-}}(\mathbf{x},\mathbf{p}), \mathbf{q}) } {\partial x_M} \\
          \vdots & \vdots & \vdots   \\
          \frac{\partial h (\phi^{-\tau_{-}}(\mathbf{x},\mathbf{p}), \mathbf{q}) } {\partial x_1}  &  \ldots  &  
          \frac{\partial h (\phi^{-\tau_{-}}(\mathbf{x},\mathbf{p}), \mathbf{q}) } {\partial x_M} \\
          \frac{\partial h (\mathbf{x}, \mathbf{q}) } {\partial x_1}  &  \ldots &  
          \frac{\partial h (\mathbf{x}, \mathbf{q}) } {\partial x_M}  \\
          \frac{\partial h (\phi^{\tau_{+}}(\mathbf{x},\mathbf{p}), \mathbf{q}) } {\partial x_1}  &  \ldots  &  
          \frac{\partial h (\phi^{\tau_{+}}(\mathbf{x},\mathbf{p}), \mathbf{q}) } {\partial x_M} \\
          \vdots & \vdots & \vdots   \\
          \frac{\partial h (\phi^{D_{+}\tau_{+}}(\mathbf{x},\mathbf{p}), \mathbf{q}) } {\partial x_1}  &  \ldots  &  
          \frac{\partial h (\phi^{D_{+}\tau_{+}}(\mathbf{x},\mathbf{p}), \mathbf{q}) } {\partial x_M} \\
               \end{array}   \right)  
\end{equation*}

\begin{equation*} 
       B = \left(    \begin{array}  {ccc}
          \frac{\partial h (\phi^{-D_{-}\tau_{-}}(\mathbf{x},\mathbf{p}), \mathbf{q}) } {\partial p_1} &  \ldots &  
          \frac{\partial h (\phi^{-D_{-}\tau_{-}}(\mathbf{x},\mathbf{p}), \mathbf{q}) } {\partial p_P}  \\
          \vdots & \vdots & \vdots \\
          \frac{\partial h (\phi^{-\tau_{-}}(\mathbf{x},\mathbf{p}), \mathbf{q}) } {\partial p_1} &  \ldots &  
          \frac{\partial h (\phi^{-\tau_{-}}(\mathbf{x},\mathbf{p}), \mathbf{q}) } {\partial p_P}  \\
          0 & \ldots & 0    \\
          \frac{\partial h (\phi^{\tau_{+}}(\mathbf{x},\mathbf{p}), \mathbf{q}) } {\partial p_1} &  \ldots &  
          \frac{\partial h (\phi^{\tau_{+}}(\mathbf{x},\mathbf{p}), \mathbf{q}) } {\partial p_P}  \\
          \vdots & \vdots & \vdots \\
          \frac{\partial h (\phi^{D_{+}\tau_{+}}(\mathbf{x},\mathbf{p}), \mathbf{q}) } {\partial p_1} &  \ldots &  
          \frac{\partial h (\phi^{D_{+}\tau_{+}}(\mathbf{x},\mathbf{p}), \mathbf{q}) } {\partial p_P}  \\
              \end{array}   \right)  
\end{equation*}

\begin{equation*} 
       C = \left(    \begin{array}  {ccc}
          \frac{\partial h (\phi^{-D_{-}\tau_{-}}(\mathbf{x},\mathbf{p}), \mathbf{q}) } {\partial q_1}  & \ldots &  
          \frac{\partial h (\phi^{-D_{-}\tau_{-}}(\mathbf{x}, \mathbf{p}), \mathbf{q}) } {\partial q_L}  \\
          \vdots & \vdots & \vdots  \\
          \frac{\partial h (\phi^{-\tau_{-}} (\mathbf{x},\mathbf{p}), \mathbf{q}) } {\partial q_1}  & \ldots &  
          \frac{\partial h (\phi^{-\tau_{-}} (\mathbf{x}, \mathbf{p}), \mathbf{q}) } {\partial q_L}   \\
          \frac{\partial h (\mathbf{x}, \mathbf{q}) } {\partial q_1}  & \ldots &  
          \frac{\partial h (\mathbf{x}, \mathbf{q}) } {\partial q_L}   \\
          \frac{\partial h (\phi^{\tau_{+}}(\mathbf{x},\mathbf{p}), \mathbf{q}) } {\partial q_1}  & \ldots &  
          \frac{\partial h (\phi^{\tau_{+}}(\mathbf{x}, \mathbf{p}), \mathbf{q}) } {\partial q_L}   \\
          \vdots & \vdots & \vdots  \\
          \frac{\partial h (\phi^{D_{+}\tau_{+}}(\mathbf{x},\mathbf{p}), \mathbf{q}) } {\partial q_1}  & \ldots &  
          \frac{\partial h (\phi^{D_{+}\tau_{+}}(\mathbf{x}, \mathbf{p}), \mathbf{q}) } {\partial q_L}  \\
                                \end{array}   \right)   . \\
\end{equation*}   

\begin{widetext}
${}$

and it can also be written as
\begin{eqnarray} \label{jacobian}
  & & DG = \\ \nonumber
  & & 
  \small{
                                    \left(    \begin{array}  {ccc}
                    \nabla_x h ( \phi^{-D_{-}\tau_{-}} (\mathbf{x},\mathbf{p}), \mathbf{q})    \cdot   D_x\phi^{-D_{-}\tau_{-}} (\mathbf{x},\mathbf{p})  
                 & \nabla_x h ( \phi^{-D_{-}\tau_{-}} (\mathbf{x},\mathbf{p}), \mathbf{q})    \cdot   D_p\phi^{-D_{-}\tau_{-}} (\mathbf{x},\mathbf{p})  
                 & \nabla_q h ( \phi^{-D_{-}\tau_{-}} (\mathbf{x},\mathbf{p}), \mathbf{q})        \\
                     \vdots  &  \vdots & \vdots \\                                    
                    \nabla_x h ( \phi^{-\tau_{-}} (\mathbf{x}, \mathbf{p}), \mathbf{q})    \cdot   D_x\phi^{-\tau_{-}} ( \mathbf{x}, \mathbf{p})  
                 & \nabla_x h ( \phi^{-\tau_{-}} (\mathbf{x}, \mathbf{p}), \mathbf{q}) )  \cdot   D_p\phi^{-\tau_{-}} ( \mathbf{x}, \mathbf{p})                  
                 & \nabla_q h ( \phi^{-\tau_{-}} (\mathbf{x},\mathbf{p}), \mathbf{q})  \\
                     \nabla_x h ( \mathbf{x}, \mathbf{q})  
                 &  0   
                 & \nabla_q h (\mathbf{x},\mathbf{q})  \\
                    \nabla_x h ( \phi^\tau_{+} (\mathbf{x}, \mathbf{p}), \mathbf{q})    \cdot   D_x\phi^\tau_{+} ( \mathbf{x}, \mathbf{p})  
                 & \nabla_x h ( \phi^\tau_{+} (\mathbf{x}, \mathbf{p}), \mathbf{q}) )    \cdot   D_p\phi^\tau_{+} ( \mathbf{x}, \mathbf{p})                  
                 & \nabla_q h ( \phi^\tau_{+} (\mathbf{x},\mathbf{p}), \mathbf{q})  \\
                      \vdots  &  \vdots & \vdots \\
                    \nabla_x h ( \phi^{D_{+}\tau_{+}} (\mathbf{x},\mathbf{p}), \mathbf{q})    \cdot   D_x\phi^{D_{+}\tau_{+}} (\mathbf{x},\mathbf{p})  
                 & \nabla_x h ( \phi^{D_{+}\tau_{+}} (\mathbf{x},\mathbf{p}), \mathbf{q})    \cdot   D_p\phi^{D_{+}\tau_{+}} (\mathbf{x},\mathbf{p})  
                 & \nabla_q h ( \phi^{D_{+}\tau_{+}} (\mathbf{x},\mathbf{p}), \mathbf{q})        \\
                                    \end{array}   \right)
                                    }
\end{eqnarray}    

\end{widetext}

where 
\begin{eqnarray} \label{gradx}
    \nabla_x h (\mathbf{x}, \mathbf{q})  & = & \left( \frac{\partial h}{ \partial x_1}, \ldots, \frac{ \partial h} {\partial x_M} \right) (\mathbf{x}, \mathbf{q})  \\
    \nabla_q h (\mathbf{x}, \mathbf{q})  & = & \left( \frac{\partial h}{ \partial q_1}, \ldots, \frac{ \partial h} {\partial q_Q} \right) (\mathbf{x}, \mathbf{q}) \label{gradq}
\end{eqnarray}
and $ D_x\phi^t ( \mathbf{x}, \mathbf{p}) $ and $D_p\phi^t ( \mathbf{x}, \mathbf{p}) $
denote the Jacobian matrices of the flow $\phi^t$ whose elements are  derivatives with respect to the state variables
$\mathbf{x}$ and the parameters $\mathbf{p}$, respectively.
For discrete dynamical systems \eqref{discrsyst2} the Jacobians $ D_x\phi^t ( \mathbf{x}, \mathbf{p}) $ and
$ D_p\phi^t ( \mathbf{x}, \mathbf{p}) $  can be computed using the chain rule and the recursion schemes
\begin{eqnarray} \label{x_recursion}
D_x\phi^{t+1} ( \mathbf{x}, \mathbf{p})  & = &    D_x g ( \phi^{t} (\mathbf{x}, \mathbf{p}), \mathbf{p}) \cdot D_x\phi^{t} ( \mathbf{x}, \mathbf{p})  \\
\nonumber              
D_p\phi^{t+1} ( \mathbf{x}, \mathbf{p})  & =  &   D_x g ( \phi^{t} (\mathbf{x}, \mathbf{p}), \mathbf{p}) \cdot D_p g(\phi^t ( \mathbf{x}, \mathbf{p}), \mathbf{p}) \\ 
\label{p_recursion}     & & +D_p\phi^{t} ( \mathbf{x}, \mathbf{p})
\end{eqnarray} 
with $D_x\phi^{0} ( \mathbf{x}, \mathbf{p}) = I_D$ ($D \times D$ unit matrix and $D_p\phi^{0} ( \mathbf{x}, \mathbf{p}) = 0$.
If backward iterations are required ($D_{-}> 0$ and $\tau_{-} > 0$) similar recursion schemes exist based on the inverse map $g^{-1}$ 
(providing $\phi^{-t}$) and its Jacobian matrice $D_x g^{-1}$ and $D_p g^{-1}$.
Instead of recursion schemes one may also use symbolic or automatic differentiation \cite{AD}.
For continuous systems \eqref{contsyst2} the required Jacobians can be obtained by 
simultaneously solving linearized systems equations as will be discussed in Sec.~\ref{ODEs}.
Inverse maps ($D_{-} > 0$ and $\tau_{-}>0$)  may be computed via backward integration of the ODEs 
(at least for short periods of time before solutions diverge significantly).
An extension for multivariate time series is straightforward.

\subsection{Parameter estimation for the H\'enon map}
\begin{figure} [!ht]
\centering
 \includegraphics[width=8.7cm]   {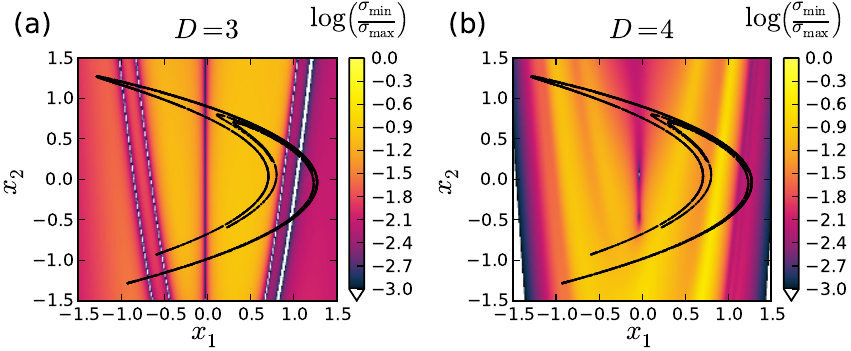}
 \caption{(Color online) Logarithm of ratio smallest singular value $\sigma_{min} = \sigma_3$  divided
by largest singular value $\sigma_{max} =\sigma_1$ vs.  $x_1$ and $x_2$ for 
the case of $M+P=3$ unknown $(x_1, x_2, p_1=a)$. The diagram
shows the plane $p_1=a=1.4$ in the three dimensional estimation space.
The other parameters are $b=0.3$ and $c=0.5$. Diagrams (a) and (b) 
show the results obtained with forward delay reconstruction dimensions $D=3$ and
$D=4$, respectively. }
\label{fig7}
\end{figure}

%
We shall now extend the discussion to include not only state estimation but also parameter estimation.
For better readability only forward delay coordinates are considered in the following, 
but all steps can also be done with mixed or backward delay coordinates, of course. 
We first consider the case where $b$  and $c$ are assumed to be known and only 
$a$ has to be estimated. In this case, $M=2$ unknown variables and $P=1$ unknown system parameter
exists (while $Q=0$). Therefore,  delay coordinates with 
dimension $D=3$ or higher will be used. 
Figure~\ref{fig7} shows the ratio of singular values $\sigma_{min}/\sigma_{max} =\sigma_3/ \sigma_1$  vs. $(x_1,x_2)$ in a 
plane in $\mathbb{R}^3$ given by $p_1 = a = 1.4$ (and fixed parameters $b=0.3$ and $c = 0.5$).
For reconstruction dimension $D=3$ the ration $\sigma_{min} / \sigma_{max}$ is very small 
for extended subsets $(\mathbf{x},\mathbf{p}) = (x_1,x_2,p_1)$ of the plane (white stripes in Fig.~\ref{fig7}a).
If the delay reconstruction dimension is increased to $D=4 $ (Fig.~\ref{fig7}b). these regions shrink or disappear.
If the dimension $D$ is increased furthermore,
the delay coordinates map is locally invertible in the full range of $x_1$ and $x_2$ values  shown in 
Fig.~\ref{fig7} (results not shown here).

\begin{figure} 
\centering
\includegraphics[width=8.7cm]    {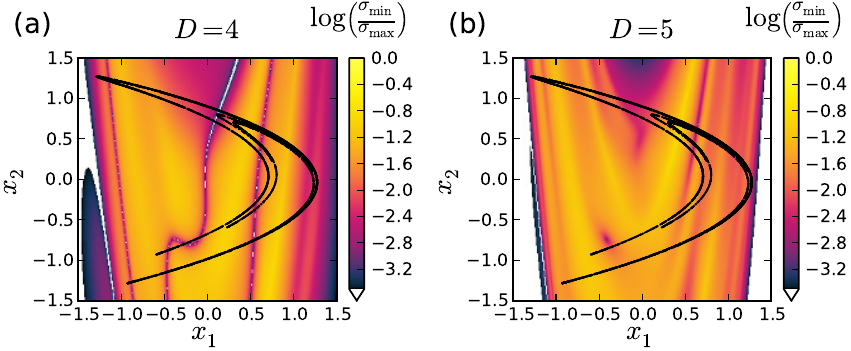}
\caption{(Color online) Logarithm of the ratio of singular values  
$\sigma_{min} / \sigma_{max} = \sigma_4 / \sigma_1$ vs.  $x_1$ and $x_2$ for 
the case of $M+P=4$ unknown quantities $(x_1, x_2, p_1=a, p_2=b)$. The diagrams
show the $x_1$-$x_2$ plane at fixed $p_1=a=1.4$ and $p_2=b = 0.3$ in the four 
dimensional estimation space for $c=0.5$. Diagrams (a) and (b) 
show the results obtained with forward delay reconstruction dimensions $D=4$ and
$D=5$, respectively. }
\label{fig8}
\end{figure}

Now we include $p_2 = b$ in the list of quantities to be estimated.
Figure~\ref{fig8} shows the ratio of singular values $\sigma_{min} / \sigma_{max}$ for
reconstruction dimensions $D=4$ and $D=5$. For $D=4$ curves with very low singular value
ratios $\sigma_{min} / \sigma_{max}$  exist crossing the H\'enon attractor which 
disappear for $D=5$.

\begin{figure} 
\centering
\includegraphics[width=8.7cm]    {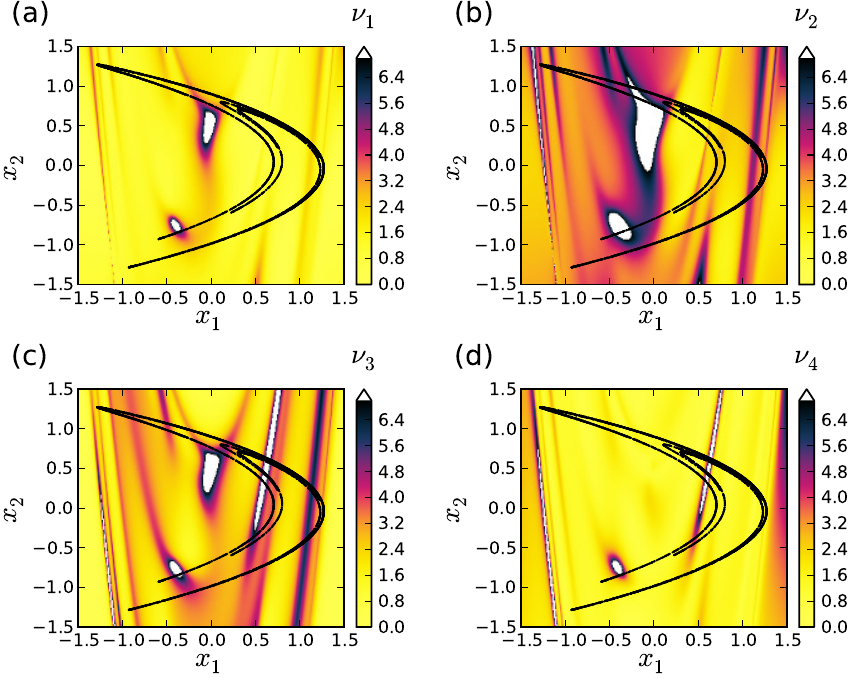}
\caption{(Color online) Estimation of uncertainties $\nu_j$ ($ j=1,\ldots,M=4)$) of variables $(x_1,x_2)$ 
and parameters $(p_1=a, p_2=b)$ 
of the H\'enon map \eqref{Henon} obtained  with $D=5$ dimensional forward delay coordinates .}
\label{fig9}
\end{figure}

Figure~\ref{fig9} shows the uncertainties $\nu_1, \ldots, \nu_4$  (Eq.~(\ref{uncert})) for $D=5$.
As can be seen, the values of uncertainties vary strongly in the $x_1$-$x_2$ plane and still
some islands with rather large uncertainties exist.

\begin{figure} 
\centering

\includegraphics[width=8.7cm]    {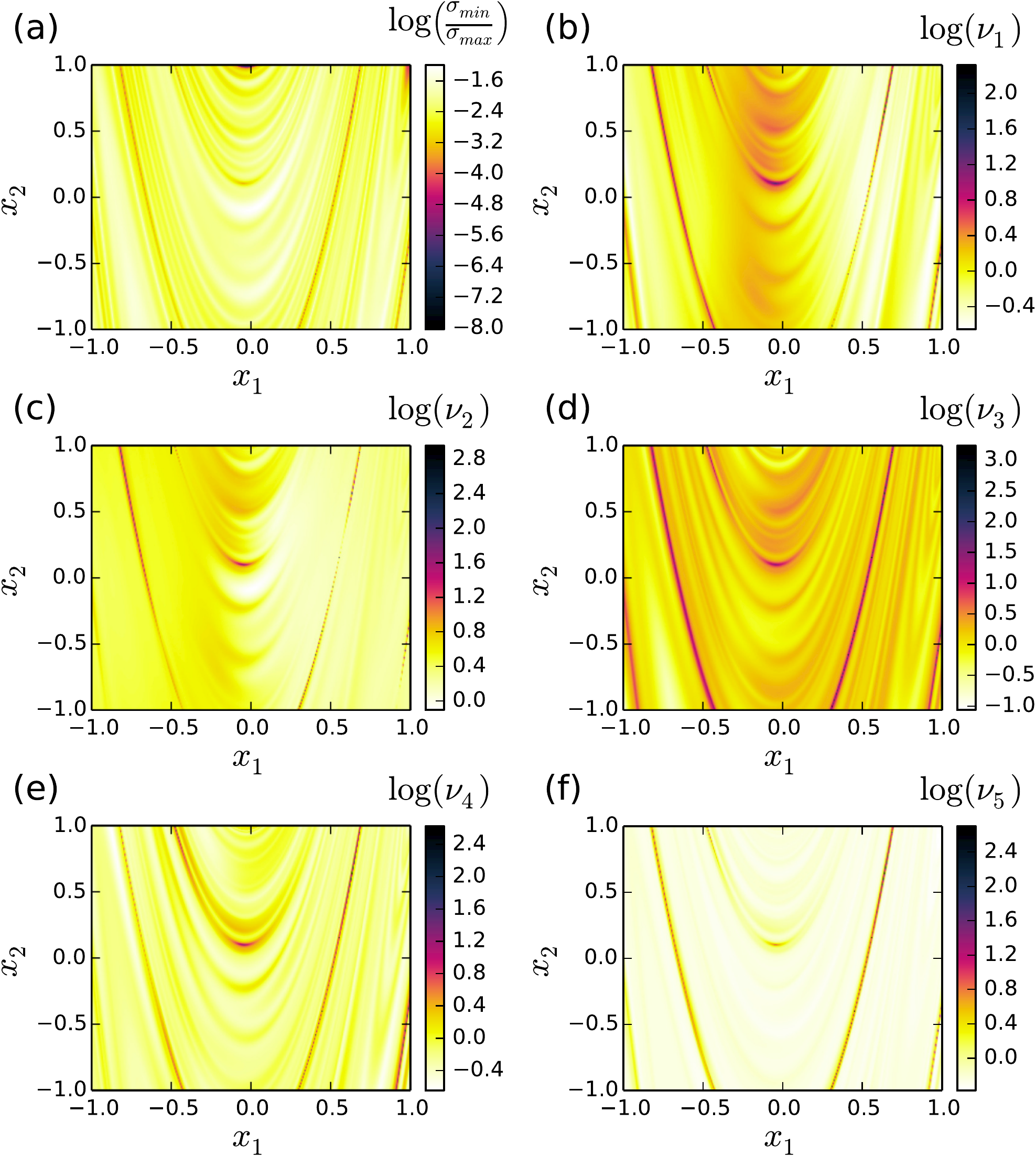}
\caption{Estimation of all variables $\mathbf{x} = (x_1,x_2)$ and 
model parameters $\mathbf{p} =(p_1,p_2) = (a,b)$ of the H\'enon map \eqref{Henon}, 
and the measurement function parameter $q=c$ (FIR filter \eqref{FIR}).
The output of the FIR filter is forward embedded in $D=9$ dimensions.
 (a) Ratio of smallest and largest singular value.
 (b)-(f) Uncertainties $\nu_j$ of state variables and parameters  in the plane
    $\{  (x_1,x_2,p_1,p_2,q):  ((x_1,x_2) \in [-1.,1.] \times [-1., 1.],
      p_1 = a = 1.4,  p_2 = b = 0.3, q = c = 0.5 \}$. }.
\label{fig10}
\end{figure}

Similar results are obtained if we include  the remaining parameters $c$ in the 
estimation problem. Scanning the two-dimensional 
$x_1$-$x_2$ subspace (plane) of the $M+P=5$ dimensional estimation problem
for $(\mathbf{x},\mathbf{p}) = (x_1, x_2, p_1 , p_2, q)$ with fixed $p_1 = a=-1.4$,
$p_2 = b =0.3$, and $q = c =0.5 $ indicates (almost) vanishing smallest
singular values as long as $D \le 8$. With $D=9$ dimensional 
delay coordinates the Jacobian matrix $DG(\mathbf{x},\mathbf{p})$ has clearly 
full rank almost everywhere within the chosen range $(x_1,x_2) \in [-1.,1.] \times [-1., 1.]$ 
as can be seen in Fig..~\ref{fig10}a.
Figures~\ref{fig10}b,c,d,e,f illustrate the  uncertainties $\nu_1, \ldots,  \nu_5$ 
(Eq.~(\ref{uncert})) of  $x_1, x_2, p_1, p_2, q $, respectively.

\subsection{Continuous dynamical systems}  \label{ODEs}
To compute the Jacobian matrix $DG({\mathbf x}, \mathbf{p})$  (\ref{jacobian}) of the delay coordinates map $G$ we 
have to compute the gradients (\ref{gradx}) and (\ref{gradq}) of the observation function $s = h(\mathbf{x}, \mathbf{q})$
and the Jacobian matrices $D_x\phi^t  (\mathbf{x},\mathbf{p})$  and $D_p\phi^t  (\mathbf{x},\mathbf{p})$
containing derivatives of the flow $\phi^t$ generated by the dynamical 
system (\ref{contsyst2}) with respect to variables $x_j$ and parameters $p_j$, respectively.
The $M \times M$-matrix  $D_x\phi^t (\mathbf{x}, \mathbf{p})$  can be computed by solving the 
linearized dynamical equations in terms of a matrix ODE
\begin{equation} \label{lineq}
  \frac{d}{dt} Y = D_xf(\phi^t(\mathbf{x}, \mathbf{p}), \mathbf{p}) \cdot Y 
\end{equation}
where $\phi^t(\mathbf{x},\mathbf{p}) $ is a solution of Eq.~(\ref{contsyst2}) with initial value $\mathbf{x}$ and
$Y$ is an $M \times M$ matrix that is initialized as $Y(0) = I_M$, where $I_M$ denotes the 
$M \times M$ identity matrix.
Similarly, the $M \times P$-matrix  $D_p\phi^t (\mathbf{x}, \mathbf{p})$ is obtained as a solution of the 
matrix ODE  \cite{Kawakami}
\begin{equation} \label{lineq2}
  \frac{d}{dt} Z = D_xf(\mathbf{x}(t), \mathbf{p}) \cdot Z +  D_pf(\mathbf{x}(t), \mathbf{p})
\end{equation}
with $Z(0)=0$.
Solving \eqref{lineq} and \eqref{lineq2} simultaneously with the system  ODEs 
(\ref{contsyst2})  we can compute $D_x\phi^\tau (\mathbf{x})$,  $D_x\phi^{2\tau} (\mathbf{x})$, etc.
and use these matrices to obtain the Jacobian matrix $DG$ of the delay coordinates map $G$ (\ref{jacobian}).
For mixed or backward delay coordinates the required components can be computed by integrating the
system ODE and the linearized ODEs backward in time.

\subsubsection{The R\"ossler system} \label{roesec}

To demonstrate the observability analysis for continuous systems
we follow Aguirre and Letellier \cite{AL05} and consider the  R\"ossler system
\begin{eqnarray}
\dot x_1 & = & - x_2 - x_3     \nonumber \\
\dot x_2 & = & x_1 + ax_2      \label{roe} \\
\dot x_3 & = & b + x_3 (x_1-c) \nonumber 
\end{eqnarray}
with $a=0.1  $, $b=0.1 $, and $c=14 $.

Time series of different observables $x_1$, or $x_2$, or $x_3$ are considered, all 
of them consisting of $N=10000$ values sampled with $\Delta t = 0.1$.
Figure~\ref{fig11} shows the R\"ossler attractor where color indicates
the uncertainty of estimating the  variable $x_1$ (first column), or $x_2$ (second column), 
or $x_3$ (third column) using forward delay coordinates. 
The results in the first row are obtained when observing $x_1$,
while the diagrams in the second and third row  show results for $x_2$ or $x_3$  time 
series. The reconstruction dimension equals $D=7$ and the delay time is $\tau = 0.5$. 
The bright yellow bullet indicates the 
state with the lowest uncertainty. This state and the $D-1 = 6$ following states plotted as
thick red bullets underly the time series values that are used for the delay reconstruction.
They span a window in time of length $(D-1) \tau = 6 \cdot 0.5 = 3$ which is about one half 
of the mean period of the chaotic oscillations $T\approx 6$.
The lowest uncertainties are obtained for states 
whose reconstruction involves trajectory segments following the vertical $x_3$-excursion
on the attractor. In contrast, trajectory segments starting from states with poor observability (large uncertainty) 
are located in the flat part of the R\"ossler attractor.
Figures~\ref{fig11}a and b show that using $x_1$ time series low values of 
$\nu_1$ occur on parts for the attractor where $\nu_2$ is high (and vice versa). Interestingly,
this is not the case for delay reconstructions based on $x_2$ time series as can be seen in 
Figs.~\ref{fig11}d and e, where low uncertainties of $x_1$ and $x_2$ occur in similar regions
on the attractor.

\begin{widetext}

\begin{figure}  [!htb]
\centering

 \includegraphics[width=17.5cm]    {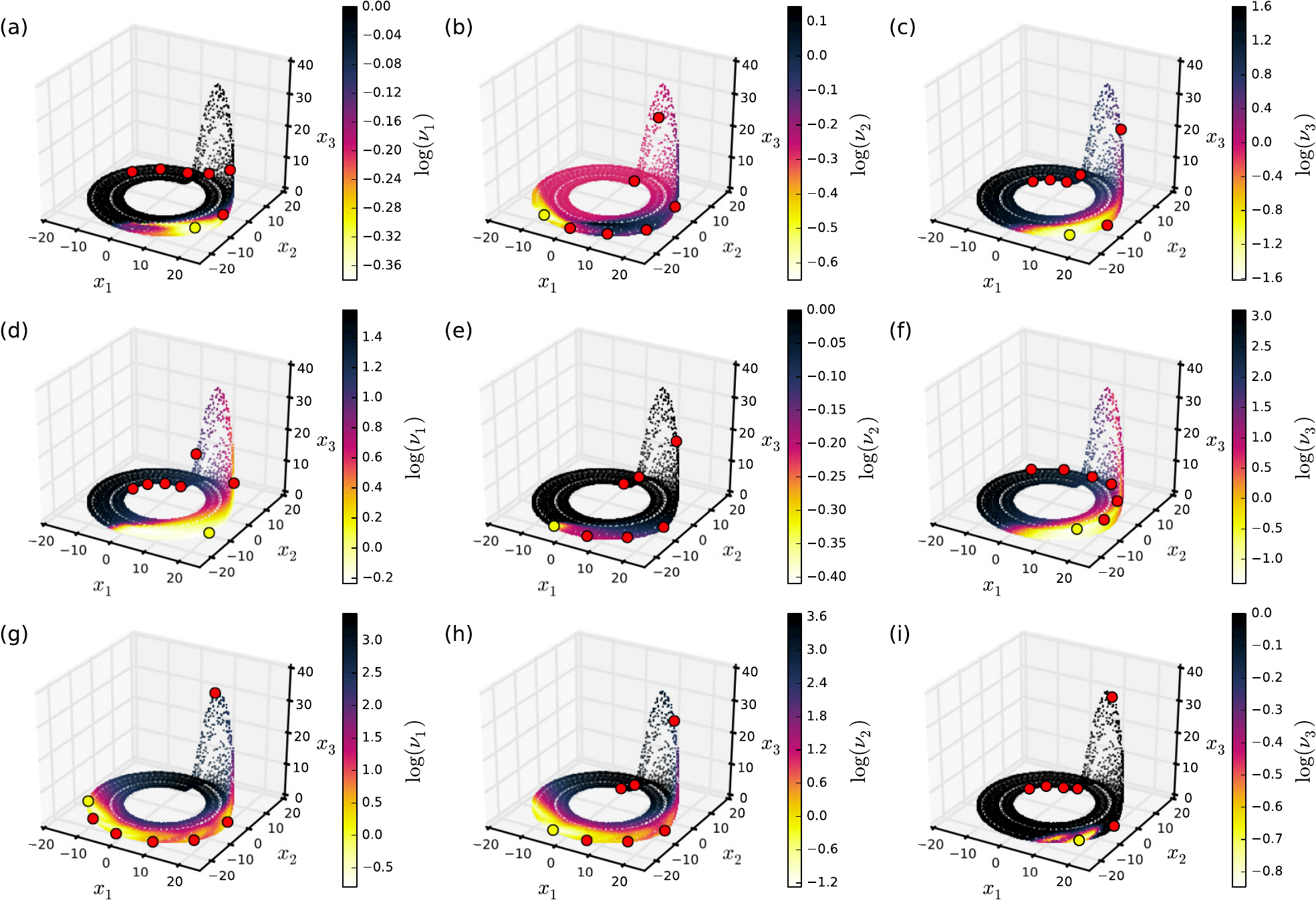}
 \caption{(color online) Color coded R\"ossler attractors where colors of points representing states
are given by logarithms of uncertainty values $\nu_1$ in the first column, $\nu_2$ in the second column,
and $\nu_3$ in the third column. All results are computed using forward delay coordinates.
The diagrams in the first row show results obtained based on an 
reconstruction of a $x_1$ time series, the second row using $x_2$ as an observable, and the third row uncertainties 
of estimates from $x_3$ data. The reconstruction dimension is $D=7$ for all nine diagrams. The bright yellow   
bullet indicates the state with the lowest $\nu$-value (respectively). To estimate this state time series 
values at this state and at $D-1 = 6$ subsequent states indicated by dark (red) bullets are used for delay coordinates.}
\label{fig11}
\end{figure}

\begin{figure} 
\centering
   
\includegraphics[width=17.cm]   {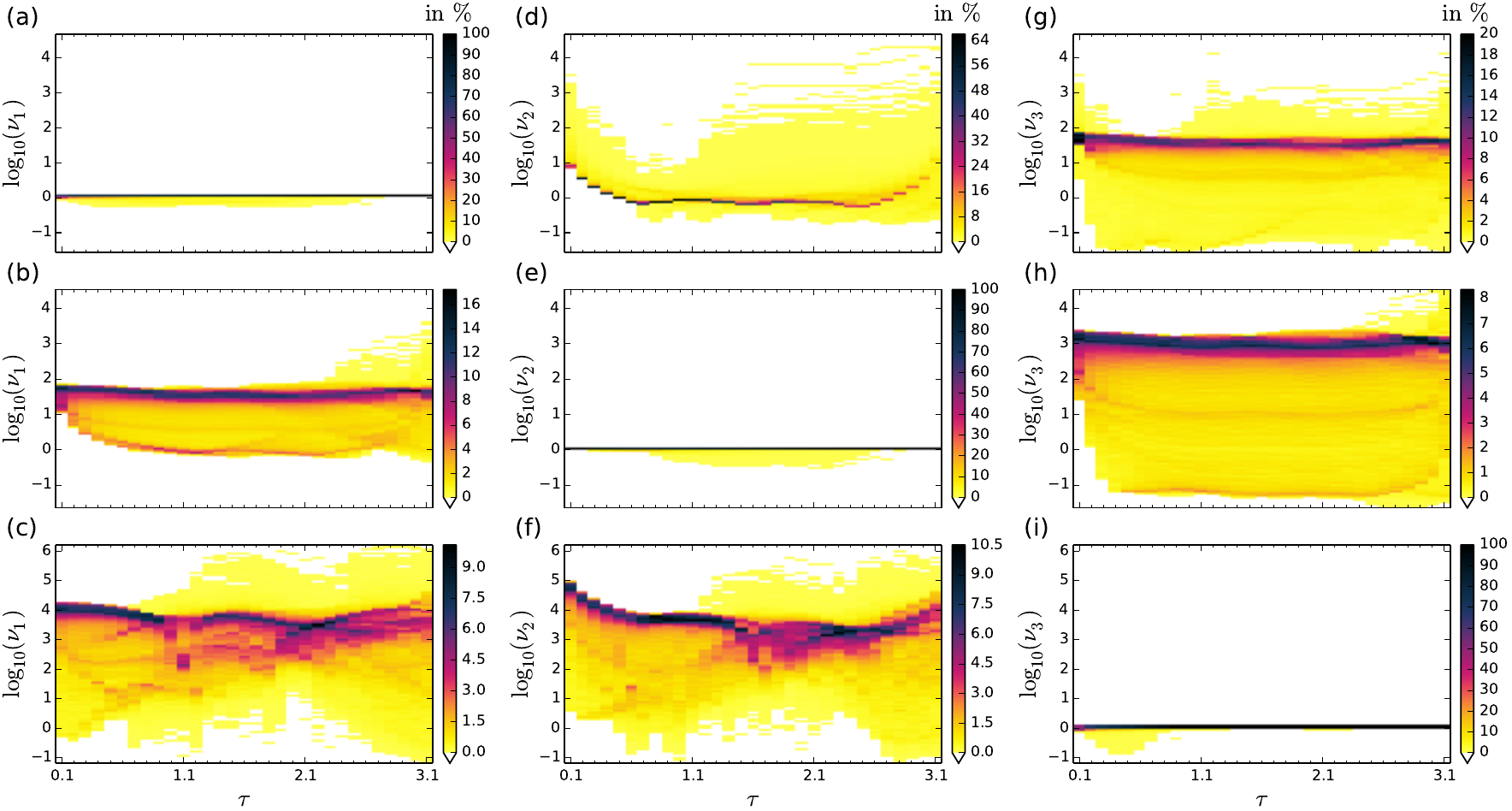}
\caption{(Color online) Histograms of uncertainties $\nu_j$ of the R\"ossler system \eqref{roe} vs.
delay time $\tau$. All parameters are assumed to be known ($M=3$, $P=0$) and forward delay coordinates
with dimension $D=4$ are used. In the first row a $x_1$ time series is given, in the second row 
$x_2$ data, and in the third row the delay reconstruction is based on $x_3$. The three columns show 
histograms of the (logarithm of the) uncertainties  $\nu_1$, $\nu_2$, and $\nu_3$, respectively.} 
\label{fig12}
\end{figure}

\begin{figure} 
\centering
   
\includegraphics[width=16.5cm]    {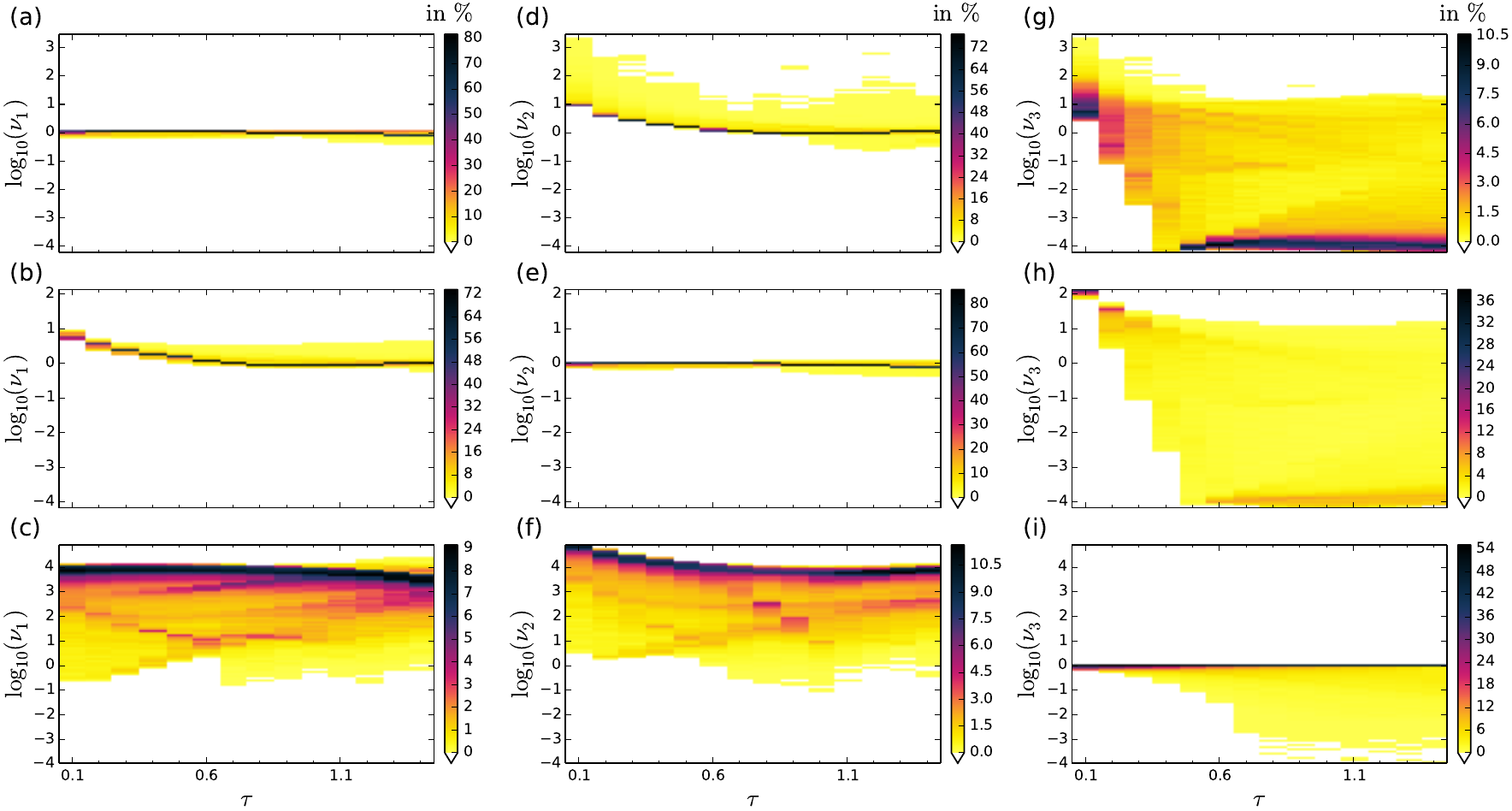}
\caption{(Color online) Histograms of uncertainties $\nu_j$ of the R\"ossler system \eqref{roe} vs.
delay time $\tau$. All parameters are assumed to be known ($M=3$, $P=0$) and mixed delay coordinates
with dimension $D=1+D_{-}+D_{+} = 4$ are used ($D_{-} = 1$, $D_{+} = 2$). 
 In the first row a $x_1$ time series is given, in the second row 
$x_2$ data, and in the third row the delay coordinates are based on $x_3$. The three columns show 
histograms of the (logarithm of the) uncertainties  $\nu_1$, $\nu_2$, and $\nu_3$, respectively.} 
\label{fig13}
\end{figure}

\end{widetext}

 In Fig.~\ref{fig12} distributions of uncertainty values of the R\"ossler system are shown that were obtained along an orbit 
 of $N=10000$ states sampled with $\Delta t = 0.1$. The distributions are shown as color coded histograms,
 estimated from the relative frequency of occurrence of the corresponding $\nu_j$ (in \%).
 All diagrams show the dependance of the histograms on the delay time $\tau$ 
 chosen for forward delay coordinates (horizontal axis). The reconstruction dimension is for all cases $D=4$ and
 all three parameters are assumed to be known (and are not part of the  estimation task, i.e. $P=0$).
 In the first row (Figs.~\ref{fig12}a,d,g) estimations are based on a $x_1$ time series from the  R\"ossler system,
 and in rows two and three, $x_2$ and $x_3$ time series are used, respectively. The uncertainties 
 $\nu_j$ of the given observable $x_j$ (Figs.~\ref{fig12}a,e,i)  mostly equal one ($\log_{10}(\nu_j) \approx 0$) or 
 are smaller (due to the additional information provided by the delay coordinates). 
 In general, lowest uncertainties for all variables are obtained when using $x_1$ time series  
 (Figs.~\ref{fig12}a,d,g) while $x_3$ data provide highest uncertainties (Figs.~\ref{fig12}c,f,i).

Figure~\ref{fig13} shows the same diagrams but now computed using four dimensional mixed delay 
coordinates with $D_{-} = 1$ and $D_{+}=2$. Similar to the results obtained with the H\'enon map the uncertainties 
computed for mixed delay coordinates are typically smaller than those obtained with forward coordinates. 
Furthermore, the histograms shown in Fig.~\ref{fig13} suggest that for mixed delay coordinates using $x_2$ as measured variable 
provides the best results, followed by the $x_1$ time series. This is in contrast to forward coordinates (Fig.~\ref{fig12}) where 
$x_1$ data yield the smallest  uncertainties for the other variables ($x_2$ and $x_3$). Similar results have been obtained 
with three dimensional forward or mixed coordinates.

\begin{figure} 
\centering
\includegraphics[width=8.6cm]    {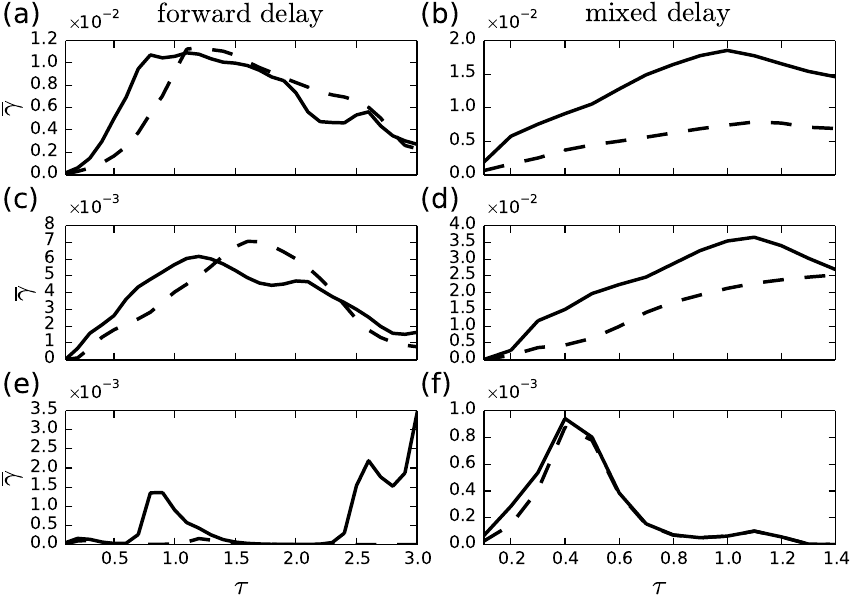}
\caption{Mean observability indices \eqref{meanobsindx} of the R\"ossler system \eqref{roe}
vs. delay time $\tau$
based on  three dimensional (dashed lines) and four dimensional (solid lines) delay coordinates. 
Left column ((a), (c), (e) forward delay coordinates. Right column ((b), (d), (f)) mixed delay coordinates with 
$D_{-}=1$ and $D_{+} = 1$ (dashed lines) or $D_{+}=2$ (solid lines). The measured time series is in the first, second, 
and third row the variable $x_1$, $x_2$, and $x_3$, respectively.    
}
\label{fig14}
\end{figure}

The fact that mixed 
delay coordinates provide the lowest uncertainties when using $x_2$ time series is consistent with results for derivative coordinates 
obtained by Letellier et al. \cite{LAM05} who found a ranking $x_2 \rhd x_1 \rhd x_3$ (for a different set of model parameters). 
For better comparison with their results  we computed the (attractor) average 
\begin{equation}  \label{meanobsindx}
  \bar  \gamma = \frac{ 1 }  { T }    \int_{0}^T \gamma(\mathbf{x}(t)) dt   
\end{equation}
of the ratio
\begin{equation}
                \gamma(\mathbf{x}) = \frac{ \sigma^2_{min} ( DG( \mathbf{x} ) ) }  {   \sigma^2_{max} ( DG( \mathbf{x} ) ) }
\end{equation}
that provides the delay reconstruction analog $ \bar  \gamma $ of the observability index \eqref{obsindx}. 
Figure~\ref{fig14} shows $\bar \gamma$ vs. the delay time $\tau$  for different delay coordinates (rows) 
and different measured  time series (columns). While for forward delay coordinates 
the largest values of the observability index occur if $x_1$ is measured (Fig.~\ref{fig14}c), $x_2$ time series 
provide best observability if mixed delay coordinates are used (Fig.~\ref{fig14}d). Note that in most cases
high observability occurs for $\tau \approx 1$ which is very close to the first zero of the autocorrelation 
function (that is often used as preferred value for delay reconstruction).

\begin{figure} 
\centering

\includegraphics[width=8.7cm]    {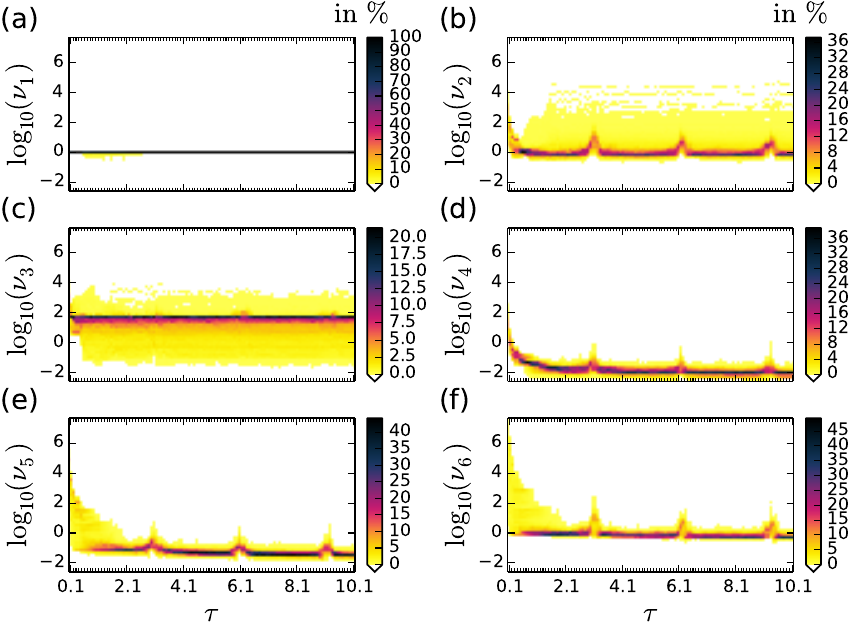}
\caption{(color online) Histograms of uncertainties $\nu_j$ of the R\"ossler system \eqref{roe} vs.
delay time $\tau$. Both, all state variables ($M=3$) and all parameters ($P=3$) are assumed to be unknown
and have to be estimated from a $x_1$ time series. 
Results are obtained using forward delay coordinates with  $D=13$.
The uncertainties $\nu_1, \nu_2, \nu_3$ correspond to state variables $x_1$, $x_2$, and $x_3$,
while $\nu_4, \nu_5, \nu_6$ quantify uncertainties of estimated parameters $p_1$, $p_2$, and $p_3$.}
\label{fig15}
\end{figure}

 Fig.~\ref{fig15} shows similar histograms but now for the full estimation problem ($M=3$ variables 
 and $P=3$ parameters). Forward delay coordinates are used and
 the reconstruction dimension is increased to $D=13$ and an $x_1$ time series
 of length $N=10000$ is used (with sampling time $\Delta t = 0.1$). For delay times $\tau$ that are an
 integer multiple of half of the mean period $T/2 \approx 3$ relatively high uncertainties of occur,
 in particular for $\nu_2$, $\nu_4$, $\nu_5$, and $\nu_6$. This is due to the well known fact that for these
 delay times the attractor reconstruction results in points scattered near a straight (diagonal) line (an effect 
 that also occurs when considering delay coordinates of a sinusoidal signal).

\section{Conclusion}
Starting from the question ``Does some particular (measured) time series provide sufficient information for 
estimating a state variable or a model parameter of interest'' we revisited the observability problem for 
nonlinear (chaotic) dynamical systems. In particular we considered delay coordinates and the ability to 
recover not measured state variables and parameters from delay vectors. This requires to ``invert'' the 
delay coordinates construction process which is at least locally possible, if the Jacobian matrix of the 
delay coordinates map has maximum (full) rank. Furthermore, we investigated how states near the 
delay vector are mapped back to the state and parameter space of the systems. In this way it is 
possible to quantify the amplification of small perturbations in delay reconstruction space in different 
directions of the state and parameter space. This reasoning gave rise to the concept of uncertainties 
of estimated variables and parameters. Both, observability and uncertainties may vary considerably 
in state space and on a given (chaotic) attractor. This feature was demonstrated with a discrete time system 
(filtered H\'enon map) and a continuous system (R\"ossler system). Local observability and uncertainties also
depend on the available measured variable (time series) and the type of delay coordinates. Best results 
were obtained with mixed delay coordinates, containing  at least a one step backward in time.

The obtained information about (local) uncertainties in state and parameter estimation can be used 
in several ways for subsequent analysis. 
First of all, it may help to decide whether the planned estimation task
is feasible at all or whether another observable has to be measured instead. For continuous time systems
relevant time scales (delay times) can be identified where uncertainties are minimal.

The strong variations of local uncertainty values in state space (along a trajectory) occurring with the examples 
shown here are typical and should be taken into account by any estimation method. If the system is in a 
state where, for example, the uncertainty $\nu_1$ of the first variable is high then it might be better not to try to 
estimate this variable in this state or close to it, because the estimate might be poor and may spoil
the overall results. Instead it makes more sense to wait until the trajectory enters a region of state space 
where $x_1$ can be estimated more reliably from the given time series. 

The concrete implementation of such an adaptive approach depends on details 
of the estimation algorithm. For Newton-like algorithms, for example, it may consist of a simple strategy 
decreasing correction step sizes. Another potential application of uncertainty analysis is the identification of 
redundant parameters, i.e., parameter combinations that provide the same dynamical output. 

\begin{acknowledgements}
The research leading to these results has received funding from the European CommunityÕs Seventh Framework 
Program FP7/2007-2013 under grant agreement no HEALTH-F2-2009-241526, EUTrigTreat. 
We acknowledge financial support by the German Federal Ministry of Education and Research (BMBF) 
Grant No. 031A147, by the Deutsche Forschungsgemeinschaft (SFB 1002: Modulatory Units in Heart Failure), 
and by the German Center for Cardiovascular Research (DZHK e.V.).
\end{acknowledgements}

\end{document}